\documentclass[11pt]{JHEP}
\usepackage{latexsym}
\usepackage{amssymb,amsfonts}
\usepackage{amsfonts}
\font\mybb=msbm10 at 11pt
\def\bb#1{\hbox{\mybb#1}}

\renewcommand{\theequation}{\arabic{section}.\arabic{equation}}
\preprint{Sept. 2, 2013; V2: minor corr.,  Oct. 30, 2013}
\title{Supersymmetric non-Abelian multiwaves in  D=3 AdS superspace}

\author{
Igor A. Bandos $^{\dagger\ddagger}$
\\
$^{\dagger}$ Department of Theoretical Physics, University of the Basque Country UPV/EHU,
\\
P.O. Box 644, 48080 Bilbao, Spain
 \\
 $^{\ddagger}$
IKERBASQUE, Basque Foundation for Science, 48011, Bilbao, Spain}

\date{September 2, 2013; minor corrections: October 30, 2013}

\abstract{We present a covariant, supersymmetric and $\kappa$--symmetric action for non-Abelian multiwave system ({\it nAmW}) in D=3  AdS superspace. Its  flat superspace limit provides a simplest counterpart of the recently proposed action for 11 dimensional system of  N nearly coincident M-waves (multiple M$0$-branes), which is presently known for the case of flat target superspace only. }


\begin{document}

\section{Introduction}

Recently a supersymmetric and $\kappa$--symmetric  action for a system of nearly coincident M-waves (or M0-branes; we call it  mM$0$--system) in flat 11D superspace has been constructed  \cite{mM0action} and studied \cite{mM0eqs}. The presence of fermions, explicit Lorentz invariance and supersymmetry can be considered as its advantage over the Myers `dielectric brane' action \cite{Myers:1999ps} and its generalization to 11D \cite{YLozano+=0207,Lozano:2005kf} \footnote{We should notice that
the supersymmetric and completely covariant action for multiple D$p$--branes was proposed in \cite{Howe+Linstrom+Linus} in
the frame of boundary fermion approach. However, this is done on the 'minus one quantization level' in the sense that, to
pass to an action similar to the action for single Dp-brane (see \cite{B+T=Dpac} and refs. therein) we have  to perform
quantization of the boundary fermion sector. Notice also the existence of the covariant and supersymmetric action for multiple D$0$--branes
system \cite{Dima+=JHEP03}, also in flat target superspace,  and an 'ectoplasma-type' action for the case of spacetime filling 3D ${\cal N}=2$ multiple D2-brane system \cite{Drummond:2002kg}. We briefly comment on these in concluding section 5. }. However, these latter are formulated in arbitrary background of the bosonic fields of
supergravity, while the action of \cite{mM0action} is presently known in the flat 11D superspace only and the construction
of its curved superspace generalization is not straightforward. To attack this problem it is desirable to have a simpler
counterparts of the mM0 system, a kind of toy models on which one might try to develop the methods to be applied then to
mM0.

Here we develop the simplest model of such a type, which can be called D=3 supersymmetric non-Abelian multiwave or, abbreviating, {\it nAmW}. We first construct the nAmW  action in flat 3D superspace and present the local 1d supersymmetry transformations which leave this action invariant. As far as our nAmW action is also invariant, by construction, under 3D target (super)space supersymmetry, it can be called doubly supersymmetric.  We also generalize this nAmW action for the case of D=3 AdS superspace, thus providing the first example of a supersymmetric multiple brane action in a curved superspace.

It should be possible to obtain the flat superspace version of our nAmW action  by dimensional reduction of the mM0 action of \cite{mM0action}. However the presence of highly constrained spinor moving frame variables make the problem of dimensional reduction of this mM0 action, as well as the study of its properties,  technically involving.

In contrast, the moving frame variables of our present D=3 system are just a pair of two component real spinors $(v_\alpha^-,v_\alpha^+)$ normalized by $v^{\alpha +}v_\alpha^-=1$, so that the derivation of equations of motion and the study of the symmetry properties might look more transparent.

Moreover, in Sec. 3 we also present our main results, the expressions for the nAmW action in D=3 AdS superspace and the supersymmetry
transformations, in terms of unconstrained spinors, without any reference to the spinor moving frame formalism. Thus the  reader who are not interested neither in this formalism nor  in a form of the action more similar to its 11D mM0 counterpart, can just read the main results in these form in Sec. 3, and omit the Sec. 4 as well as sec. 2.3 and Appendix, where the spinor
moving frame formalism is used and discussed. However, in our opinion the structure of  the nAmW action is much more transparent in its spinor moving frame form, which also shows similarity to the 11D mM$0$ system.

This paper is organized as follows. In Sec. 2 we review the twistor-like formulations of 3D massless superparticle in flat and curved $D=3$ superspaces; sec. 2.3. discusses the spinor moving frame reformulation of this twistor like superparticle action (see \cite{B90,IB+AN=95} and refs therein).
In Sec. 3 we resume our results by writing  the D=3 nAmW action and its worldline supersymmetry ($\kappa$--symmetry)  transformations in AdS superspaces in terms of unconstrained spinors. In Sec. 4, to clarify the geometric structure of the nAmW action and its relation
with 11D mM$0$ action of \cite{mM0action}, we reformulate the D=3 nAmW action in terms of spinor moving frame
variables. Sec. 4.1. is devoted to the spinor moving frame form of the  nAmW action  in flat superspace. In Sec. 4.2. we discuss the geometrical meaning of different terms of this action and its relation with the action for 1d reduction of d=2 supersymmetric Yang--Mills (SYM) theory. The relation with 11D mM$0$ action of \cite{mM0action} is briefly discussed in sec. 4.3. In sec. 4.4. we obtain  the generalization of the
nAmW action for the case of D=3 AdS superspace. We conclude by outlook in Sec. 5. The Appendix contains some more  details on spinor moving frame formalism.

\section{Twistor--like formulation of the massless superparticle in  D=3 ${\cal N}$=1 superspaces}

\subsection{Massless superparticle in flat 3D superspace}

The action of D=3 massless superparticle can be written in the  form (see \cite{Ferber:1977qx,Shirafuji:1983zd} for D=4
counterpart as well as \cite{Dima99} and refs. therein)
\begin{eqnarray}
\label{S0=3D} S^{3D}_0 &=& \int_{W^1}  {\cal L}_1^0 = \int_{W^1}  \lambda_\alpha\lambda_\beta \hat{E}^a
\tilde{\gamma}_a^{\alpha\beta}\; , \quad
\end{eqnarray}
where $\lambda_\alpha (\tau)$ is an auxiliary bosonic
spinor field and  \begin{eqnarray} \label{hEa=3D}\hat{E}^a =E^a (\hat{Z}(\tau)) =d\tau \hat{E}^a_\tau \, , \qquad \hat{E}^a_\tau =
(\partial_\tau \hat{x}^a- i\partial_\tau\hat{\theta} \gamma^a \hat{\theta} )\; ,
\end{eqnarray}
is the pull--back to the worldline of the bosonic supervielbein form of the D=3 ${\cal N}=1$ flat superspace, \footnote{Our conventions for $D=3$ Dirac matrices are
\begin{eqnarray}
\label{3dgamma}
 \gamma^a\gamma^b = \eta^{ab}+ i\epsilon^{abc}\gamma_c\; , \qquad \gamma^a_{\alpha\beta}
 = -i \gamma^a_\alpha{}^\sigma\epsilon_{\sigma\beta}\; , \qquad \tilde{\gamma}{}^{a\alpha\beta}
= i \epsilon^{\alpha\sigma}\gamma^a{}_\sigma{}^{\beta} \; , \qquad a=0,1,2\; , \qquad \nonumber  \\ \nonumber \gamma^a\tilde{\gamma}^b+
\gamma^b\tilde{\gamma}^a
 = 2\eta^{ab}\; , \qquad  \epsilon^{\alpha\beta}= \left(\matrix{0& 1 \cr -1 & 0}\right)= - \epsilon_{\alpha\beta}\; , \qquad  \epsilon^{\alpha\sigma} \epsilon_{\sigma\beta}
 = \delta^\alpha_\beta\; .
\qquad
\end{eqnarray}}
\begin{eqnarray}\label{Ea=3D}
 E^a:= E^a(Z)= dx^a - id\theta \gamma^a \theta\; .
\end{eqnarray}
This is to say $\hat{E}^a$ of (\ref{hEa=3D}) is obtained from $E^a(Z)$ of (\ref{Ea=3D})  by substituting the coordinate functions   $\hat{Z}^M(\tau)= (\hat{x}^a(\tau), \hat{\theta}^\alpha (\tau))$ for the superspace  coordinates ${Z}^M= ({x}^a, {\theta}^\alpha)$ in (\ref{Ea=3D}). These coordinate functions are used to define the superparticle worldline $W^1$ as a line in superspace,
\begin{eqnarray}
\label{W1inSSP} W^1\in \Sigma^{(3|2)}\; : \qquad {Z}^M= \hat{Z}^M(\tau)\; .
  \end{eqnarray}

Due to the famous $D=3$($,4,6,10$) identity $\gamma_{a\alpha (\beta}\gamma^a{}_{\gamma\delta)}=0$, the vector
$\lambda\gamma^a\lambda:= \lambda^\alpha \gamma^a_{\alpha\beta}\lambda^\beta \equiv \lambda_\alpha
\gamma^{a\alpha\beta}\lambda_\beta $ is light--like, $\lambda\gamma^a\lambda\, \lambda\gamma_a\lambda\equiv 0$. This
indicates the relation of the twistor--like (Ferber--Shirafuji type \cite{Ferber:1977qx,Shirafuji:1983zd})  action
(\ref{S0=3D}) for $D=3$($,4,6,10$) massless superparticle with the Brink--Schwarz action \cite{Brink:1981nb}. This latter can be written in the first order
form
\begin{eqnarray}
\label{SBS=3D} S_{BS} &=& \int_{W^1} (p_a \hat{E}^a - d\tau {e\over 2}p_ap^a) \; , \quad
\end{eqnarray}
the variation of which with respect to Lagrange multiplier $e=e(\tau)$ produces the mass shell constraint $p_ap^a\approx 0$. In D=3,
solving this algebraic constraint by
\begin{eqnarray}
\label{p=vgv=3D} p_a = \lambda\gamma^a\lambda \; , \quad
\end{eqnarray}
 and substituting this solution back to (\ref{SBS=3D}) we arrive at (\ref{S0=3D}).

A particular convenience of the twistor--like formulation of the  massless superparticle is that in its frame the local
fermionic $\kappa$--symmetry \cite{kappaAL,kappaS} acquires irreducible form, while in the Brink--Schwarz formulation it is
infinitely reducible. Namely, the action (\ref{S0=3D}) is invariant under the transformations
\begin{eqnarray}
\label{kap0=irr} && \delta_\kappa \hat{x}^a =  - i  \hat{\theta} \Gamma^a\delta_\kappa\hat{\theta}\; , \qquad \delta_\kappa
\hat{\theta}^\alpha = \kappa (\tau)  \lambda^{\alpha} \; , \qquad \delta_\kappa \lambda^{\alpha}=0
 \; .  \quad
\end{eqnarray}
with an arbitrary fermionic function $\kappa (\tau)$.
In contrast, the  $\kappa$--symmetry of the Brink--Schwarz superparticle action  \cite{kappaS}
\begin{eqnarray}
\label{kap=red} && \delta_\kappa \hat{x}^a =  - i  \hat{\theta} \Gamma^a\delta_\kappa\hat{\theta}\; , \qquad \delta_\kappa
\hat{\theta}^\alpha =
 p^a\tilde{\gamma}_a^{\alpha\beta} \kappa_\beta  \; , \qquad  \delta_\kappa e=4i \hat{\theta}^\alpha\kappa_\alpha \; ,
 \qquad
\delta_\kappa p_a=0
 \;   \quad
\end{eqnarray}
is infinitely reducible. It is parametrized by a two component spinorial function $ \kappa_\beta= \kappa_\beta(\tau)$ but only one (of the linear combinations) of these enters the transformations of $\hat{\theta}$ efficiently. (Notice that, in contrast, the $\kappa$--symmetry of the massive ${\cal N}=2$ superparticle \cite{kappaAL} is irreducible and allows a covariant gauge fixing). We will not comment more on this well known fact
but rather stress that, in its irreducible form  the $\kappa$--symmetry can be easily identified with the local worldline supersymmetry \cite{stv} (see also \cite{Dima99}).

The local supersymmetry of our  non-Abelian multiwave action will act by a deformation of the irreducible
$\kappa$--symmetry on its center of energy variables.

\subsection{Massless superparticle in 3D AdS superspace}

%$AdS^{(3|2)}$

The action (\ref{S0=3D}) describes  also the movement of superparticle in curved superspace provided
\begin{eqnarray} \label{hEa=3Dc}\hat{E}^a =E^a (\hat{Z}(\tau)) =d\tau \hat{E}^a_\tau \, , \qquad \hat{E}^a_\tau = \partial_\tau  \hat{Z}^M(\tau) E_M^a (\hat{Z}) \; ,
\end{eqnarray}
is the pull--back to the bosonic worldline of the supervielbein of the curved D=3 superspace
\begin{eqnarray}
\label{Ea=3Dc}
 E^a:= E^a(Z)= dZ^M E_M^a(Z) \; , \qquad  E^\alpha:= E^\alpha(Z)= dZ^M E_M^\alpha (Z) \; , \qquad
  \end{eqnarray}
which obey the superspace supergravity constraints (see e.g. \cite{1001}).

In particular the action for massless superparticle   in the D=3 AdS superspace  is given by Eq. (\ref{S0=3D}) where $\hat{E}^a$ is the pull--back of the bosonic supervielbein which, together with its fermionic partner, obeys the following structure equations (Maurer--Cartan equations)\footnote{In our notation the exterior derivative $d$ acts from the right,
$dE^a= d(dZ^ME_M^a(Z))=dZ^M \wedge dZ^N \partial_N E_M^a$, where $\wedge$ denotes the exterior product of differential forms characterized by   $$E^a  \wedge E^b=- E^b  \wedge E^a\; , \qquad E^a  \wedge E^{\beta}=- E^{\beta}  \wedge E^a\; , \qquad E^{\alpha} \wedge E^{\beta}=+ E^{\beta}  \wedge E^{\alpha}\; . $$
}
\begin{eqnarray}
\label{dEa=AdS} dE^a= -i E^\alpha\wedge E^\beta \gamma^a_{\alpha\beta} + {\zeta \over 2}E^c\wedge E^b \epsilon_{bc}{}^a\; ,
\qquad dE^\alpha =  {i\zeta \over 2}E^c\wedge E^\beta \gamma_{c\beta}{}^{\alpha} \; .  \qquad
\end{eqnarray}
Here the  constant $\zeta $ has dimension of $L^{-1}$. It is proportional to inverse 'radius' of the $AdS_3$ space, the bosonic body of the  AdS superspace,
\begin{eqnarray}
\label{dimZeta=-2}
{} [\zeta ]= L^{-1}\; , \qquad \zeta \propto {1\over R_{AdS}}  \; . \qquad
  \end{eqnarray}
The irreducible $\kappa$ symmetry of the massless superparticle in curved superspace can be written in the form of
\begin{eqnarray}
\label{kap0=irrC} && \delta_\kappa \hat{Z}^M = \kappa (\tau)  \lambda^{\alpha} E_{\alpha}^M (\hat{Z})
\; , \qquad \delta_\kappa \lambda^{\alpha}=0
 \; .   \quad
\end{eqnarray}

Below, to simplify equations we will omit the hat symbols from the coordinate functions and pull--backs of the forms, so that from now on
\begin{eqnarray}
\label{hat=out}
\hat{Z}^M(\tau) \mapsto {Z}^M(\tau)\; , \qquad \hat{E}^a \mapsto E^a\; , \qquad  etc.
 \quad
\end{eqnarray}

\subsection{Spinor moving frame reformulation of the 3D massless superparticle action}

Let us introduce  a St\"uckelberg field $\rho^{\#}$ and use it to write the bosonic spinor $\lambda_\alpha$  as
\begin{eqnarray}
\label{l=vr=s2}
 \lambda_\alpha= v^-_\alpha \sqrt{\rho^{\#}}\, , \quad
\end{eqnarray}
where $v^-_\alpha( \tau)$ is another (also nonvanishing) bosonic spinor variable.
Such a composite  $\lambda_\alpha$ is invariant under  the local scalings
\begin{eqnarray}
\label{v->vs=s2} v^-_\alpha\mapsto e^{-\beta (\tau)} v^-_\alpha\; , \qquad \rho^{\#} \mapsto e^{2\beta (\tau)} \rho^{\#} \,   \quad
\end{eqnarray}
which we will identify below with the $SO(1,1)$ gauge transformations.

Substituting (\ref{l=vr=s2}) for $\lambda$ in (\ref{S0=3D}) we arrive at the following equivalent form of the twistor--like action describing the {\it spinor moving frame} formulation of the massless superparticle,
\begin{eqnarray}
\label{S0=3D=s2} S^{3D}_0 &=& \int_{W^1} \rho^{\#} \, v^{-} \gamma_{a}v^{-}\,  {E}^a  = \qquad \\
&=& \int_{W^1} \rho^{\#} {E}^{=} = \int_{W^1} \rho^{\#} {E}^a u_a^{=} \; . \quad \nonumber
\end{eqnarray}
In the second line of (\ref{S0=3D=s2}) we have introduced the light--like vector constructed as a bilinear of the bosonic spinors $v_\alpha^-$,
\begin{eqnarray}
\label{u--=3D=s2} u_a^{=}= v^-\gamma_av^- \;
\end{eqnarray}
(see Sec. 4 and Appendix for their meaning)  and its contraction with the pull--back of the supervielbein, ${E}^{=} :={E}^a u_a^{=}$.  The objects of this type will be very useful to clarify the geometrical meaning of different terms of the non--Abelian multiwave action.  We will turn to this in Sec. 4. But first, in the next Sec. 3, we will describe this action in D=3 AdS superspace  using just the unconstrained spinors.

To conclude this section, let us present the irreducible $\kappa$--symmetry of 3D massless superparticle in its spinor moving frame formulation,
\begin{eqnarray}
\label{susy-th0} \delta_\epsilon  {Z}^M = \epsilon^{+} v^{-\alpha} E_{\alpha}^M ({Z}) \; , \qquad \label{susy-x0}   \delta_\epsilon v_\alpha^\pm=0\; .
\end{eqnarray}
Notice that the fermionic parametric function $\epsilon^{+}:=\epsilon^{+}(\tau)$ in (\ref{susy-th0} ) carries the $SO(1,1)$ weight 1  and has dimension of square root of length, $[\epsilon^{+}]=L^{1/2}$, which is the natural dimension of a supersymmetry parameter.

\section{D=3, ${\cal N}$=1  supersymmetric non-Abelian multiwave in flat and AdS superspace. Summary in terms of unconstrained variables}

3D non-Abelian multiwave (3D {\it nAmW}), the $D=3$ counterpart of the multiple M0-brane action of \cite{mM0action}, is a multiparticle (multiwave) system with special type of interaction between constituents.  To describe it we need to introduce, besides the usual coordinate functions (\ref{W1inSSP}) (see also (\ref{hat=out})), describing now the center of energy motion, also a set of traceless $N\times N$ matric fields describing the relative motion of $N$ nAmW constituents.
These are a Hermitean bosonic $N\times N$ matrix field $\check{{\bb X}}(\tau)$, a hermitian fermionic
$N\times N$ matrix field $\check{\Psi}(\tau)$, as well as auxiliary variables\footnote{We put the check symbol $\check{...}$ over the
matrix variable of this section to distinguish them from the redefined fields of the next section, see (\ref{bX=rhobX++}) for their relations.}. The set of these includes the  bosonic traceless $N\times N$ matrix $\check{{\bb P}}$, playing the role of momentum for the matrix field $\check{{\bb X}}$, and $SU(N)$ connection ${\bb A}=d\tau {\bb A}_\tau (\tau)$. We also need  one more real valued (not matrix) auxiliary field. This is a bosonic spinor $w^\alpha$ subject to a topological type restriction $w^\alpha\lambda_\alpha\not=0$. It can be used to construct a spinor $\tilde{\lambda}_\alpha$ complementary to $\lambda^\alpha$,
\begin{eqnarray}
\label{tl:=} \tilde{\lambda}_\alpha= {w_\alpha \over \lambda w}\; , \qquad \lambda w:= \lambda^\beta w_\beta
 \; , \qquad \lambda^\alpha\tilde{\lambda}_\alpha = 1
 \;  .  \quad
\end{eqnarray}

The Lagrangian form of the action for 3D non-Abelian multiwave in  ${\cal N}=1$, $D=3$ AdS superspace
\begin{eqnarray}
\label{SnAW:=}
 && S^{AdS}_{nAmW} =\int_{W^1} {\cal L}_1^{ nAw} = \int_{W^1} ({\cal L}_1^{0} + {1\over \mu^6} {\cal L}_1^{SU(N)})  \; , \qquad
 {\cal L}_1^{ nAw} = {\cal L}_1^{0} + {1\over \mu^6} {\cal L}_1^{SU(N)}\; , \qquad
\end{eqnarray}
can be split on the center of energy Lagrangian
\begin{eqnarray}
  \label{L0:=} && {\cal L}_1^{0}= \lambda\gamma_a\lambda {E}^{a}\; ,
\end{eqnarray}
which has the same form as in the case of single massless superparticle, and the  Lagrangian form for the matrix fields describing relative motion. It is denoted by $ {\cal L}_1^{SU(N)}$, as far as our matrices
are traceless and hence take values in $su(N)$ algebra. Before describing the  form of $ {\cal L}_1^{SU(N)}$, let us notice that it enters in the action (\ref{SnAW:=}) divided by 6th power of parameter $\mu$ which  has the dimension of mass (this is to say the  length dimension $(-1)$, $[\mu ]=L^{-1}$, as far as in our notation  $\hbar=1$, $c=1$).
The dimensions of our matrix variables are
 \begin{eqnarray}
\label{dimX=-2} [\check{{\bb X}}]= L^{-2}\; , \qquad [\check{\Psi}]= L^{-3}\; , \qquad  [\check{{\bb P}}]= L^{-4}\; . \qquad
  \end{eqnarray}
This seemingly strange dimensions allow to minimize the number of appearances of the dimensional parameter $\mu$ in the Lagrangian and in the supersymmetry transformations leaving it invariant (see  below).

 The Lagrangian 1--form $ {\cal L}_1^{SU(N)}$ in (\ref{SnAW:=}) reads  \footnote{See the next section for the relation of ${\cal L}_1^{SU(N)}$ to 1d reduction of 2d SYM Lagrangian.}
 \begin{eqnarray}
\label{LSUN:=}  {\cal L}_1^{SU(N)}  &=& - tr\left(\check{{\bb P}} D \check{{\bb X}}\right) + {i\over 4} tr \left(\check{\Psi} D
\check{ \Psi}  \right)+  \tilde{\lambda}\gamma_a\tilde{\lambda} {E}^{a}  \check{\cal H} + d{\theta}^{\alpha}
\tilde{\lambda}_{\alpha} tr\left( \check{\Psi}\check{{\bb P}} \right)+ d{\lambda}^{\alpha} \tilde{\lambda}_{\alpha} tr\left(
\check{{\bb P}} \check{{\bb X}} \right) - \quad \nonumber \\ && -  {\zeta \over \mu^6}\int_{W^1}  \tilde{\lambda}^\alpha
d\tilde{\lambda}_\alpha \, (tr(\check{{\bb P}}\check{{\bb X}}))^2
\; .   \quad
\end{eqnarray}
where
\begin{eqnarray}
\label{cH=} \check{\cal H}    = {1\over 2} tr\left(\check{{\bb P}} \check{{\bb P}} \right)  - 2  tr\left(\check{{\bb X}}
\check{\Psi}\check{\Psi}\right)   \;  . \qquad
  \end{eqnarray}
As it will be clear after reformulating the action in spinor moving frame formalism, $\check{\cal H}  $ has a meaning of the Hamiltonian of the relative motion and its origin can be traced to the Hamiltonian of the 1d reduction of d=2 $SU(N)$ SYM model.
The covariant derivatives
\begin{eqnarray}
\label{DX=} D\check{{\bb X}} :=d\tau D_\tau \check{{\bb X}}  = d\check{{\bb X}} + [{\bb A},    \check{{\bb X}}] \; , \qquad
 D\check{\Psi} = d\check{\Psi}+ [{\bb A},  \check{\Psi} ] \;  \qquad
  \end{eqnarray}
include the $SU(N)$ connection  on $W^1$, ${\bb A}= d\tau {\bb A}_\tau (\tau)$ were
${\bb A}_\tau (\tau)$ is an anti-Hermitian traceless $N\times N$ matrix gauge field in 1d, which is an independent variable in our model.

The dimensional parameter  $\zeta$ (\ref{dimZeta=-2}), which entered the structure equations of AdS superspace (\ref{dEa=AdS}),  is present in the last term of Eq. (\ref{LSUN:=}) only.  Notice the second appearance of ${1\over \mu^6}$ multiplier which is necessary to equate the dimansion of these and other terms in ${\cal L}_1^{SU(N)} $.
Setting $\zeta=0$, which  implies omitting the last terms and using the flat superspace supervielbein $E^\alpha=d\theta^\alpha$ and (\ref{Ea=3D})\footnote{The structure equations (\ref{dEa=AdS}) with $\zeta=0$ describe flat superspace and can be solved by (\ref{Ea=3D}) and $E^\alpha=d\theta^\alpha$.}, we arrive at the nAmW  action in flat superspace.

The action (\ref{SnAW:=}) is invariant under the 3D supersymmetry of AdS superspace by construction.
Furthermore, it  is invariant under the local, $\tau$--dependent  supersymmetry transformations which act on the relative motion variables by
\begin{eqnarray}
\label{susyX=}
 {}  \delta_\epsilon \check{{\bb X}}= i \epsilon\check{\Psi}\; , \qquad \delta \check{{\bb P}}=0\; , \qquad \delta \check{\Psi}=
 2\epsilon \check{{\bb P}} \; , \qquad
{}  \delta_\epsilon {\bb A}= i {E}^{a} \tilde{\lambda}\gamma_a\tilde{\lambda} \;   \epsilon\check{\Psi} + 16
{E}^{\alpha} \tilde{\lambda}_{\alpha} \, \epsilon  \check{{\bb X}}
 \; ,  \quad
\end{eqnarray}
and on the center of energy variables as a deformation of the irreducible $\kappa$--symmetry transformations (\ref{kap0=irrC})
\begin{eqnarray}
\label{kap=irrZ} && \delta_\kappa {Z}^M = \kappa (\tau)  \lambda^{\alpha} E_{\alpha}^M ({Z}) + {3i\over 4\mu^6 }
\tilde{\lambda}\gamma^a\tilde{\lambda}\, tr (\epsilon \check{\Psi}\check{{\bb P}}) E_{a}^M ({Z})
\; , \qquad
\\
 \label{kap=irr+l-}
&& \delta_\epsilon  \lambda_{\alpha}= -i {\zeta\mu^6} \epsilon\, (tr(\check{{\bb P}}\check{\Psi})) \,\tilde{\lambda}_{\alpha} \; ,
\qquad \delta_\epsilon  \tilde{\lambda}_{\alpha}=0
 \; .  \quad
\end{eqnarray}
Notice that the worldline  supersymmetry `parameter' $\epsilon(\tau)$ in (\ref{kap=irrZ}), (\ref{kap=irr+l-}) has the dimension of length, $[\epsilon  ]=L$, like the irreducible $\kappa$--symmetry `parameter' in (\ref{kap0=irr}).

In the case of flat superspace the worldline supersymmetry transformations of the center of energy variables read
\begin{eqnarray}
\label{kap=irr+} && \delta_\epsilon {x}^a =  - i  {\theta} \Gamma^a\delta_\epsilon {\theta}+ {3i\over 4\mu^6 }
\tilde{\lambda}\gamma_a\tilde{\lambda}\, tr (\epsilon \check{\Psi}\check{{\bb P}}) \; , \qquad \delta_\epsilon
{\theta}^\alpha = \epsilon  (\tau)  \lambda^{\alpha} \; , \qquad \\ \label{kap=irr+l} && \qquad \delta_\epsilon
\lambda^{\alpha}=0 \; , \qquad \delta_\epsilon  \tilde{\lambda}_{\alpha}=0
 \; . \quad
\end{eqnarray}
One can easily appreciate that in this case only the variation of the bosonic coordinate function $\delta_\epsilon {x}^a$ is deformed (with respect to (\ref{kap0=irr})) by the contribution of the matrix fields. In the case of AdS superspace the deformations touches also the bosonic spinor $\lambda_\alpha$ which, in distinction to the case of flat superspace, is not inert under the supersymmetry, Eq. (\ref{kap=irr+l-}).

\section{Spinor moving frame formulation of the 3D non-Abelian multiwave action in flat and AdS superspaces}

\label{SmfSec}

To make clear the geometric structure of the nAmW action and its relation with the 11D multiple M$0$ action of
\cite{mM0action,mM0eqs}, in this section we reformulate it using the D=3 spinor moving frame variables. We will begin from the action in flat D=3 superspace and then, after comparison with 11D mM$0$ action, pass to its AdS generalization.

\subsection{Spinor moving frame action  of the 3D non-Abelian multiwave in flat superspace and its local worldline supersymmetry }

After a field redefinition, the flat superspace limit ($\zeta=0$) of the 3D non-Abelian multiwave action (\ref{SnAW:=}) can be presented in a more compact form
\begin{eqnarray}
\label{SmM0=3D} S_{nAmW} &=& \int\limits_{W^1} \rho^{\#}{E}^{=} + {1\over \mu^6} \int\limits_{W^1} (\rho^{\#})^3 \left(
tr\left(- {\bb P}{\cal D} {\bb X} + {i\over 4} { \Psi} {\cal D} {\Psi}  \right) + {E}^{\#} {\cal H} + i {E}^{+}
tr({\Psi}  {\bb P})\right) , \qquad
\\
\label{HSYM=3D} && {\cal H}=   {1\over 2} tr\left( {\bb P} {\bb P} \right)  - 2\,  tr\left({\bb X}\, \Psi{\Psi}\right) \;
\qquad
  \end{eqnarray}
which is also more similar to the multiple M-wave action of \cite{mM0action,mM0eqs}\footnote{Notice that, although  the
counterpart of the dimensional parameter $\mu$ is set equal to unity in \cite{mM0action,mM0eqs}, one can easily restore it
in any equation of \cite{mM0action,mM0eqs} by balancing the dimensions of left and right hand sides. }.

Here $\rho^{\#}$  is a St\"uckelberg field (compensator) for $SO(1,1)$ symmetry (see (\ref{v->vs=s2}) and (\ref{v->vs}) below), which is one of the gauge symmetries of the action (\ref{SmM0=3D}). It has been already introduced in Eq. (\ref{l=vr=s2}) which, as we will see below, is also used in our field redefinition. The matrix fields in the  action (\ref{SmM0=3D}) carries nontrivial weights with respect to the $SO(1,1)$ gauge  symmetry, so that in a more explicit notation
\begin{eqnarray}
\label{bX=bX++} && {\bb  X}= {\bb  X}_{\#}:= {\bb  X}_{++}\, ,\qquad {\bb  P}= {\bb  P}_{\#\#}\, ,\qquad    {{ \Psi}}=
\Psi_{\# \,+}:= \Psi_{++ \,+}= \Psi_{\#}{}^-\,  .\quad \qquad
 \end{eqnarray}
Notice that  the upper $^-$ and lower $_+$ indices denote the same SO(1,1) weight, and that
the  $SO(1,1)$ weight of any of the matrix fields in (\ref{SmM0=3D}) (which is counted by upper sign indices, so that the weight of ${\bb X}$ is -2)  is equal to its doubled  dimension\footnote{The same is true for the Lagrange multiplier (St\"uckelberg field ) $\rho^{\#}$, but not for the moving frame variables which we will describe below.},
 \begin{eqnarray}
\label{dimX=-1} [{\bb X}]= L^{-1}\; , \qquad [{\Psi}]= L^{-3/2}\; , \qquad  [{\bb P}]= L^{-2}\; . \qquad
  \end{eqnarray}
In its turn these scaling dimensions  reflect the  origin of the matrix fields in the (1d reduction of the 2d) supersymmetric Yang--Mills (SYM) model. We discuss this in sec. \ref{fromSYM} below.

The $SO(1,1)$ invariant matrix fields of the previous section may be identified with the appropriate  'weighted' matrix field of this section multiplied by certain  powers of $\rho^{\#}$, {\it i.e.}
\begin{eqnarray}
\label{bX=rhobX++} \check{\bb  X} = \rho^{\#} {\bb  X} \equiv  \rho^{\#} {\bb  X}_{\#}\, ,\qquad \check{\bb  P}=
(\rho^{\#})^2 {\bb  P}\equiv (\rho^{\#})^2 {\bb  P}_{\#\#}\, ,\qquad  \nonumber \\   \check{\Psi} = (\rho^{\#})^{3/2} \Psi \equiv
(\rho^{\#})^{3/2} \Psi_{\#}^-\,  .\qquad
 \end{eqnarray}
As far as $[\rho^{\#}]=L^{-1}$, one easily reproduces the dimensions of the matrix fields in  (\ref{dimX=-2}) from
(\ref{dimX=-1}).

The $SO(1,1)$ symmetry  acts also on the bosonic and fermionic 1-forms
\begin{eqnarray}\label{E==Eu=}
{E}^{=}:= {E}^a u_{ {a}}^{=}\; ,  \qquad \label{E-=Ev-}
{E}^{-}:= {E}^\alpha v_{ {\alpha}}^{-}\; ,  \qquad
\\ \label{E++=Eu++} {E}^{\#}:= {E}^a u_{ {a}}^{\#}\; ,  \qquad
  {E}^{+}:= {E}^\alpha v_{ {\alpha}}^{+}\; .  \qquad
\end{eqnarray}
These are constructed from the pull--backs to the center of energy worldline of the supervielbein forms $E^a$ (\ref{Ea=3D})
and $E^\alpha =d\theta^\alpha$,
\begin{eqnarray}
\label{hEal=3D}{E}^a =d{x}^a- id{\theta} \gamma^a {\theta}\; , \qquad {E}^\alpha =d{\theta}^\alpha\;
\qquad
\end{eqnarray}
using the pair of bosonic spinors $v_{ {\alpha}}^{\mp}$ normalized so that
\begin{eqnarray}\label{v-v+=1}
v^{\alpha -} v_{ {\alpha}}^{+}:=\epsilon^{\alpha\beta} v_{ {\alpha}}^{+}v_{ {\beta}}^{-} =1\; .  \qquad
\end{eqnarray}
The vectors $u_{ {a}}^{=}$ and  $u_{ {a}}^{\#}$ are constructed as  bilinear of these normalized bosonic spinors,
\begin{eqnarray}
\label{u--=3D} u_a^{=}= v^-\gamma_av^- \qquad \Rightarrow \qquad 2 v^-_\alpha v^-_\beta = u_a^{=} \gamma^a_{\alpha\beta} \;
, \qquad u^{a=} u_a^{=}=0\; , \qquad
\\
\label{u++=3D} u_a^{\#}= v^+\gamma_av^+ \qquad \Leftrightarrow \qquad 2 v^+_\alpha v^+_\beta = u_a^{\#}
\gamma^a_{\alpha\beta} \; , \qquad u^{a\#} u_a^{\#}=0\; . \qquad
\end{eqnarray}
They are light-like (as has been indicated in (\ref{u--=3D}) and (\ref{u++=3D})) and normalized by $u^{a=} u_a^{\#}=2$. The third bilinear of the bosonic spinors
\begin{eqnarray}
\label{u2=3D} u_a^{\perp}= v^-\gamma_av^+ \qquad \Rightarrow \qquad v^-_\alpha v^+_\beta + v^+_\alpha v^-_\beta =
u_a^{\perp} \gamma^a_{\alpha\beta} \; , \qquad u^{a\perp} u_a^{\perp}=-1\; , \qquad
\end{eqnarray}
is spacelike and orthogonal to both $u_a^{=}$ and  $u_a^{\#}$. Resuming the properties of these composite vectors, (see \cite{Sok})
\begin{eqnarray}
\label{uu=3D}
 u^{a\#} u_a^{\#}=0 ,   \qquad u^{a=} u_a^{=}=0  ,   \qquad u^{a\#} u_a^{=}=2 ,   \qquad \nonumber \\ u^{a\#} u_a^{\perp}= 0
 ,   \qquad u^{a=} u_a^{\perp}= 0 ,   \qquad u^{a\perp} u_a^{\perp}=-1 ,   \qquad
\end{eqnarray}
we can see that they describe a moving frame attached to the center of energy worldline $W^1$  of the non--Abelian multiwave. \footnote{
Eqs. (\ref{uu=3D})  imply that our composite vectors  form an $SO(1,2)$ valued matrix,
\begin{eqnarray}
\label{UinSO=3D}
 U_a^{(b)}:= \left({1\over 2}(u_a^{\#}+u_a^{=}) ,  u_a^{\perp} , {1\over 2}(u_a^{\#}-u_a^{=}) \right)   \; \in \; SO(1,2) \nonumber
 \\ \nonumber \quad \Leftrightarrow \; U^T\eta U=\eta =diag (1,-1,-1)\, ,  \quad
\end{eqnarray}
called {\it moving frame matrix}.  Correspondingly the bosonic spinors $v_\alpha^{\pm}$ are called {\it spinor moving frame
variables} (see \cite{B90,IB+AN=95} as well as \cite{mM0action} and refs therein).}

As we have already noticed, the bosonic spinor which enters the massless superparticle action (\ref{S0=3D}) can be expressed as in Eq. (\ref{l=vr=s2}), \begin{eqnarray}
\label{l=vr}
 \lambda_\alpha= v^-_\alpha \sqrt{\rho^{\#}}\,  .  \quad
\end{eqnarray}
Its dual spinor entering the non-Abelian multiwave action (\ref{SnAW:=}) reads
\begin{eqnarray}
\label{tl=vr}
 \tilde{\lambda}_\alpha= v^+_\alpha /\sqrt{\rho^{\#}}\,  .  \quad
\end{eqnarray}
Both are inert under the $SO(1,1)$ St\"uckelberg gauge symmetry
\begin{eqnarray}
\label{v->vs} v^-_\alpha\mapsto e^{-\beta} v^-_\alpha\; , \qquad  v^+_\alpha\mapsto e^{\beta} v^+_\alpha\; , \qquad
\rho^{\#} \mapsto e^{2\beta} \rho^{\#} \,   \quad
\end{eqnarray}
which is useful for clarifying the group theoretical structure of the spinor moving frame variables. Using this symmetry as identification relation we can consider the spinors $v_\alpha^\pm$ as homogeneous coordinates of the coset $SO(1,2)/SO(1,1)$\footnote{To be more precise, the coset is $SL(2,{\bb R})/SO(1,1)=Spin(1,2)/SO(1,1)$. One easily see that, due to constraint (\ref{v-v+=1}), determinant of the 2$\times $2 matrix $V_\alpha^{(\beta )}=(v_\alpha^+,v_\alpha^-)$ is equal to unity  which is tantamount to saying that this {\it spinor moving frame matrix} belongs to $SL(2,{\bb R})$ group.  }.

The covariant derivatives in (\ref{SmM0=3D})
\begin{eqnarray}
\label{DX:=} {\cal D} {\bb X}=  d{\bb X} + 2\Omega^{(0)} {\bb X} + [\, {\bb A}\, , \, {\bb X} \,] \; , \qquad {\cal D}
{\Psi}=d{\Psi}+ 3\Omega^{(0)} {\Psi} + [\, {\bb A}\, , \, {\Psi} \,]   ,  \; \;   \qquad
  \end{eqnarray}
involves, besides the $SU(N)$ connection ${\bb A}= d\tau {\bb A}_\tau$, also $SO(1,1)$ connection $\Omega^{(0)}$ constructed  from the bosonic spinors
\begin{eqnarray}
\label{3D:Om0=} \Omega^{(0)}= v^{\alpha -}\, dv^+_\alpha = {1\over 4} u^{=a}du_a^{\#}\; . \qquad
  \end{eqnarray}
Due to the presence of this composite $SO(1,1)$ connection, the action (\ref{SmM0=3D}) is invariant under the $SO(1,1)$ gauge symmetry\footnote{In its form (\ref{SnAW:=}) the action contains the $SU(N)$ covariant derivatives (\ref{DX:=}) while the $\propto \Omega^{(0)}$ terms of (\ref{DX=}) give rise to the pre-last term in  (\ref{SnAW:=}). (Notice that
$tr (\Psi {\cal D}\Psi )= tr (\Psi {D}\Psi )$ as far as $tr \{\Psi , \Psi\}\equiv 0$).   }.  The traceless anti-hermitian $N\times N$ matrix gauge field $ {\bb A}_\tau$ is an independent dynamical  variable of our model\footnote{As it is usual for the time component of gauge field, it is not dynamical; it plays the role of the  Lagrange multiplier for the Gauss constraint. }.

The local supersymmetry transformations which leave the action (\ref{SmM0=3D}) invariant read
\begin{eqnarray}
\label{susy-th}
 \delta_\kappa {Z}^M &= & \epsilon^{+}  v^{-\alpha} E_{\alpha}^M ({Z})  +  {3i\over 4\mu^6 }  \,
\epsilon^{+} \; tr(\Psi {\bb P}  )\;   (\rho^{\#})^2 u^{a\#}  E_{a}^M ({Z})  \; , \qquad
\\
\label{susy-rho}  &&  \delta_\epsilon \rho^{\#} = 0\; , \qquad \label{susy-v}
 \delta_\epsilon  v_{\alpha}^{\pm} =0
  \quad  \Rightarrow  \quad  \delta_\epsilon
  u_a^{=}= \delta_\epsilon u_a^{\#}= \delta_\epsilon u_a^{\perp}=0\;
 ,  \qquad
\\
\label{susy-X}  \delta_\epsilon {\bb X}   &=& i \epsilon^{+} \Psi \; , \quad \delta_\epsilon {\bb P}   = 0\; ,\qquad
\label{susy-Psi}  \delta_\epsilon \Psi =  2\epsilon^{+} {\bb P}\; ,\qquad \\ \label{susy-A}
 && \delta_\epsilon {\bb A} = -  4{E}^{\#} \epsilon^{+}  \Psi + 16 {E}^{+}
 \epsilon^{+}\;    {\bb X}
 \; .  \qquad
\end{eqnarray}
Notice that the local supersymmetry parameter in the spinor moving frame formulation has its natural dimension of square
root of length,
\begin{eqnarray}
\label{dimEp=1/2} [\epsilon^{+}]= L^{1/2} \; .  \qquad
\end{eqnarray}

Although the action (\ref{SmM0=3D}) is equivalent to (\ref{SnAW:=}), which can be obtained from it by field redefinition, the structure of (\ref{SmM0=3D}) is much more transparent. Below we will show that it can be treated as 1d reduction of the 2d $SU(N)$ supersymmetric Yang--Mills (SYM) action coupled to 1d supergravity induced by the embedding of the nAmW center of energy worldline into the target superspace.

\subsection{Clarifying the structure of the non-Abelian multiwave action}

\label{fromSYM}

Let us write  the simplified version of the first two terms in the second integral in (\ref{SmM0=3D}) obtained by replacing $E^\#$ bt $d\tau$,
\begin{eqnarray}
\label{LSYM=3D}
{\cal L}^{SYM}_1=
tr\left(- {\bb P}{\cal D} {\bb X} + {i\over 4} { \Psi} {\cal D} {\Psi}  \right) + d\tau  {\cal H} , \;
  \end{eqnarray}
We observe that this is just the Lagrangian form of the  the 1d reduction of the  2d ${\cal N}=1$ supersymmetric Yang--Mills (SYM) action written in the first order formalism, and that
${\cal H}$ defined in (\ref{HSYM=3D}) is just the Hamiltonian of this 1d  SYM model\footnote{In the second order formalism, this is to say after substituting the algebraic equations of motion of the auxiliary matrix field ${\bb P}$, the Lagrangian form (\ref{LSYM=3D}) reads {\small $tr\left(-{1\over 2} {\cal D}_\tau {\bb X}{\cal D}_\tau {\bb X} + {i\over 4} { \Psi} {\cal D} {\Psi}  -  {\bb X}\{ { \Psi}, {\Psi}\}  \right)$}. From the 2d SYM perspective ${\bb X}$ originates in special component of the gauge potential, ${\bb X} ={\bb A}_1$, and ${\cal D}_\tau {\bb X} $ is the YM field strength  ${\bb F}_{01}= \partial_0 {\bb A}_1 - \partial_1 {\bb A}_0 + [{\bb A}_0, {\bb A}_1] $ calculated in the assumption that gauge potential is independent on  the second, special coordinate:  ${\bb F}_{01}=  {\cal D}_\tau{\bb X}- \partial_1 {\bb A}_0\; \mapsto \; {\cal D}_\tau{\bb X}$ when   $\partial_1 {\bb A}_0=0$.} (hence the superscript SYM of ${\cal L}^{SYM}_1$). $\int_{W^1} {\cal L}^{SYM}_1$ is invariant under the rigid supersymmetry transformations
\begin{eqnarray}
\label{susy0-X}  \delta_{\epsilon_0} {\bb X}   &=& i \epsilon_0 \Psi \; , \qquad \delta_{\epsilon_0} {\bb P}   = 0\; ,\qquad
\label{susy0-Psi}  \delta_{\epsilon_0} \Psi =  2{\epsilon_0} {\bb P}\; ,\qquad \\ \label{susy0-A}
 && \delta_{\epsilon_0}{\bb A}_\tau = -  4 {\epsilon_0}\Psi
 \;  \qquad
\end{eqnarray}
with constant fermionic parameter ${\epsilon_0}$. The conserved current generating this supersymmetry is ${\cal S}= tr (\Psi {\bb P})$. Actually one of the ways to obtain (\ref{SmM0=3D}) is to search for the locally supersymmetric generalization of the 1d SYM action with the Lagrangian form (\ref{LSYM=3D}).

The standard way to make the action invariant under {\it local} supersymmetry is to couple it to supergravity. Let us denote the 1d graviton and gravitino one forms by
${E}^{\#}=d\tau {E}_\tau^{\#}(\tau)$ and ${E}^{+}=d\tau {E}_\tau^{+}(\tau)$ (the appearance of the sign indices will be clear in a moment). Under the local supersymmetry  with parameter $\epsilon^+(\tau)$ these  should  transform   as
\begin{eqnarray}\label{v1dSG=}
\delta_\epsilon \hat{E}^{+}= D \epsilon^{+}(\tau) \; ,  \qquad \delta_\epsilon
\hat{E}^{\#}= -2i \hat{E}^{+}\epsilon^{+}\; .
\end{eqnarray}
Notice that this is exactly the transformation rules of the one-forms (\ref{E++=Eu++}) under the local worldline supersymmetry (\ref{susy-th0}) of the massless superparticle action.

This suggests to try to provide the local  supersymmetry by introducing the coupling to {\it induced} supergravity (\ref{E++=Eu++}) and to describe the dynamics of this later by adding to the SYM term $(\propto ) {\cal L}_1^{SYM}$ the Lagrangian form ${\cal L}_1^0=\rho^{\#} E^=$ formally coincident with the one of the massless superparticle action (\ref{S0=3D=s2}).

The coupling to the induced 1d graviton  is reached by replacing in (\ref{LSYM=3D}) $d\tau$ by $E^\#$. The third contribution  in the brackets of (\ref{SmM0=3D}), $iE^+ tr (\Psi {\bb P})$,  is the standard coupling of gravitino to the supercurrent of rigid supersymmetry, $iE^+{\cal S}$.   Then, to match the SO(1,1) weights and dimensions  different  terms in the modified (\ref{LSYM=3D}), one should assume the weights and dimensions of the fields as in (\ref{bX=bX++}) and (\ref{dimX=-1}), and also multiply all but one terms in the Lagrangian form  by
${(\rho^{\#})^3\over \mu^6}$. The necessity of additional dimensional constant $\mu$ becomes transparent on this stage. In such a way, starting from the 1d dimensional reduction of the 1d SYM  Lagrangian form ${\cal L}^{SYM}$ (\ref{LSYM=3D}), we have arrived at $ {(\rho^{\#})^3\over \mu^6} \int\left( {\cal L}^{SYM}_1\vert_{d\tau \mapsto E^\#}+ iE^+ tr (\Psi {\bb P}) \right) + \int {\cal L}^{0}_1$ which coincides with the nAmW action
(\ref{SmM0=3D}).

\subsection{Relation with 11D multiple M-wave system}

In the moving frame formulation of the nAmW action, (\ref{SmM0=3D}) the relation with the action for multiple M-waves (muliple mM$0$--branes or mM$0$ system) of \cite{mM0action,mM0eqs} becomes transparent. This latter  also involves the  contractions of pull--backs of bosonic supervielbein $E^{\underline{a}}$ (now eleven dimensional, with
${\underline{a}}=0,1,...,10$) with light--like moving frame (11-)vectors, $E^==E^{\underline{a}} u_{\underline{a}}^=$, $E^\# =E^{\underline{a}} u_{\underline{a}}^\#$, the latter having the meaning of the 'induced 1d graviton',   as well as  the induced gravitino 1-form  $\; E^{+q}$ with $q=1,...,16$.  This latter is given by the contraction of the pull--back of 11D fermionic supervielbein $E^{\underline{\alpha}}$ (${\underline{\alpha}}=1,...,32$) with a square root of the  light--like vector $u_{\underline{a}}^\#$, which is a highly constrained set of 16 11D spinors $v_{\underline{\alpha}}^{+q}$ (see \cite{mM0action,mM0eqs} for details),  $\; E^{+q}=E^{\underline{\alpha}} v_{\underline{\alpha}}{}^{+q}$. The action of \cite{mM0action} also contains 16 fermionic matrix fields
${\Psi}_q$ and a nanoplets of the bosonic matrix fields ${\bb X}^i$ ($i=1,...,9$) and their auxiliary 'momenta' ${\bb P}^i$ carrying the same $SO(1,1)$ weights and dimensions as our singlets $\Psi$, ${\bb X}$ and ${\bb P}$. Without more explanations and details let us write the formal expression for  the mM$0$ action in flat 11D superspace \cite{mM0action}\footnote{The dimensional constant $\mu$ is set to unity in \cite{mM0action,mM0eqs}.},
\begin{eqnarray}
\label{SmM0=}  S_{mM0} &=& \int_{W^1} \rho^{\#}\, {E}^{=}
+ \int_{W^1} {(\rho^{\#})^3\over \mu^6}\, \left(  tr\left(- {\bb P}^i {\cal D} {\bb X}^i + 4i { \Psi}_q{\cal D}
{\Psi}_q  \right) + {E}^{\#} {\cal H} \right)+ \quad \nonumber \\ &&  +  \int_{W^1}
 {(\rho^{\#})^3\over \mu^6}\, {E}^{+q}  tr\left(4i (\gamma^i {\Psi})_q  {\bb P}^i + {1\over 2}
(\gamma^{ij} {\Psi})_q  [{\bb X}^i, {\bb X}^j]  \right) , \;
 \\
\label{HmM0=} {\cal H} &=&  {1\over 2} tr\left( {\bb P}^i {\bb P}^i \right)  - {1\over 64}
tr\left[ {\bb X}^i ,{\bb X}^j \right]^2  - 2\,  tr\left({\bb X}^i\, \Psi\gamma^i {\Psi}\right) .  \qquad
  \end{eqnarray}
Here $\gamma^i=\gamma^i_{pq}$ are $SO(9)$ Dirac matrices and $\gamma^{ij}= \gamma^{[i}\gamma^{j]}$.

In addition to the straightforward generalizations of the terms of (\ref{SmM0=3D}), this mM$0$ action conatins the potential term $ {\cal V}  = - {1\over 64} tr\left[ {\bb X}^i ,{\bb X}^j \right]^2 $ in (\ref{HmM0=}) and the Yukawa coupling $\propto  {1\over 2}
 E^{+q} tr\left(
(\gamma^{ij} {\Psi})_q  [{\bb X}^i, {\bb X}^j]\right)$ in the first line of (\ref{SmM0=}), both constructed with the use of commutator $[{\bb X}^i, {\bb X}^j]$  of the bosonic matrix field. This commutator vanishes if we restrict ourselves to the configuration with only one bosonic matrix field ${\bb X}^i \mapsto  {\bb X}$ (say, ${\bb X}^i = {\bb X}\delta^{i 1}$)  which is certainly assumed by reduction of the 11D  mM$0$ system to a D=3 multiple brane model. The true reduction of the fermionic matrix fields, $\Psi_q\mapsto \Psi$, and of the center of energy sector, $(x^{\underline{a}}, \theta^{\underline{\alpha}}) \mapsto (x^{{a}}, \theta^{{\alpha}})$ and $ v_{\underline{\alpha}}{}^{\pm q}\mapsto  v_{{\alpha}}{}^{\pm}$ requires more discussion, and this is beyond the score of this paper.

The above schematic consideration already allows us to recognize  in (\ref{SmM0=3D}) the 3D counterpart of (\ref{SmM0=}). Also one can appreciate that, despite of relative simplicity of the original massless superparticle action (\ref{S0=3D}) in comparison with (\ref{S0=3D=s2}), the nAmW action (\ref{SnAW:=}) written explicitly in terms of unconstrained spinors is not so simple and its structure looks obscure. Furthermore,
it is not so easy to find the similarity between (\ref{SnAW:=})  and the 11D spinor moving frame action  (\ref{SmM0=}), while, as we have seen above,  it is quite straightforward in the case of spinor moving frame action  (\ref{SmM0=3D}), which also has a more transparent geometric structure.\footnote{Furthermore, the only form of the mM$0$ action which is known presently is the spinor moving frame action (\ref{SmM0=}). The same is actually true concerning the twistor--like action for 11D massless superparticle \cite{IB07:M0}, the 11D counterpart of (\ref{S0=3D=s2}): its form in terms of a single unconstrained spinor is not known and is actually prohibited (at least if Lorentz covariant actions are concerned) by the properties of 11D Dirac matrices).}

\subsection{Non-Abelian multiwave in $AdS_3$ superspace}

%Spinor moving frame formulation.}

The next stage in our program is to search for a generalization of our model for the case of curved  ${\cal N}=1$ $D=3$
target superspace. It is natural to begin with the AdS superspace in which $E^a$ and $E^\alpha$ are Cartan forms
characterized by the Maurer--Cartan equations (\ref{dEa=AdS}).

If we apply literally the local supersymmetry transformation rules (\ref{susy-th})--(\ref{susy-A}) to the action
(\ref{SmM0=3D}) with $E^a$ and $E^\alpha$ obeying  (\ref{dEa=AdS}), we find the following nonvanishing  variation of the action
\begin{eqnarray}
\label{susyS=AdS} \delta_\varepsilon S_{nAmW}= {i\zeta \over \mu^6}\int_{W^1} \varepsilon^+\, (\rho^{\#})^3 {E}^\perp tr
({\bb P}\Psi ) \; . \qquad
\end{eqnarray}
 Here
\begin{eqnarray}
\label{susyS=AdS}
{E}^\perp = {E}^a u_a^\perp\; , \qquad
\end{eqnarray}
is the projection of the pull--back of the bosonic supervielbein of target superspace on the third moving frame vector    $u_a^\perp$ (see  (\ref{uu=3D}) and  (\ref{E==Eu=}), (\ref{E++=Eu++})).

This can be compensated by
\begin{itemize}
\item Allowing the spinor moving frame variables to transform under supersymmetry with
 \\ $\delta_\varepsilon v_\alpha^- = -i \zeta (\rho^{\#})^2 \varepsilon^{+}\, (tr({\bb P}{\Psi}))
\, v_\alpha^+ $, while $\delta_\varepsilon v_\alpha^+ = 0$,
\item and adding to the action $S_{nAmW}$ the additional term
 $- {\zeta  \over \mu^6} (\rho^{\#})^5 v^{\alpha +}d v_{\alpha}^{+}\, (tr({\bb P}{\bb X}))^2$.
    \end{itemize}

At this stage it is convenient to introduce covariant Cartan forms
\begin{eqnarray}
\label{f=:=} f^{=}= v^{\alpha -}d v_{\alpha}^{-} \; , \qquad \\ \label{f++:=} f^{\#}= v^{\alpha +}d v_{\alpha}^{+} \; . \qquad
  \end{eqnarray}
and to write the above described  action of non-Abelian multiwaves in $D=3$ AdS superspace as
\begin{eqnarray}
\label{SmM0=S+AdS} \fbox{$\;  S^{AdS}_{nAmW} = S_{nAmW} -  {\zeta  \over \mu^{{12}}}\int_{W^1}   (\rho^{\#})^5 f^{\#}\, (tr({\bb
P}{\bb X}))^2\;$} , \qquad  \end{eqnarray}
where the  bosonic and fermionic supervielbein forms entering the   $S_{nAmW} $ through $E^\#, E^+$ (\ref{E++=Eu++}) and $E^=$ (\ref{E==Eu=}), obey the structure equations of the AdS superspace, Eqs. (\ref{dEa=AdS}).

The action (\ref{SmM0=S+AdS}) is invariant under the local worldline supersymmetry
\begin{eqnarray}
\label{susy-ZAdS} \delta_\kappa {Z}^M &= &
\epsilon^{+}  v^{-\alpha} E_{\alpha}^M ({Z}) - {3i\over 4\mu^6 }  \, \epsilon^{+} \; tr(\Psi {\bb P}  )\;   (\rho^{\#})^2 u^{a\#}  E_{a}^M ({Z})  \; , \qquad
\\
\label{susy-rhoAdS}  &&  \delta_\epsilon \rho^{\#} = 0\; , \qquad \label{susy-v+AdS} \delta_\epsilon  v_{\alpha}^{+} =0 \quad
\Rightarrow  \quad  \delta_\epsilon
  u_a^{\#}=0\, , \qquad  \\ \label{susy-v-AdS}
\fbox{$\; \delta_\epsilon  v_{\alpha}^{-} = i_\epsilon f^{=} v_\alpha^+ \; $} &&
  \quad  \Rightarrow  \qquad  \delta_\epsilon
  u_a^{=}= u_a^{\perp}  i_\epsilon f^{=}\; , \qquad  \delta_\epsilon u_a^{\perp} = {1\over 2}i_\epsilon f^{=} u_a^{\#}\;
 ,  \qquad \\ \label{susy-if=} && \fbox{$\; i_\epsilon f^{=} =  - i {\zeta\over \mu^6} (\rho^{\#})^2 \varepsilon^{+}\, (tr({\bb
 P}{\Psi}))\; $}\; \qquad
\\
\label{susy-XAdS}  \delta_\epsilon {\bb X}   = i \epsilon^{+} \Psi \; &,& \quad \delta_\epsilon {\bb P}   = 0\; ,\qquad
\label{susy-PsiAdS}  \delta_\epsilon \Psi =  2\epsilon^{+} {\bb P}\; ,\qquad \\ \label{susy-AAdS}
 && \delta_\epsilon {\bb A} = -  4{E}^{\#} \epsilon^{+}  \Psi + 16 {E}^{+}
 \epsilon^{+}\;    {\bb X}
 \;   \qquad
\end{eqnarray}
which differs from its flat superspace version (\ref{susy-th})--(\ref{susy-A}) by nontrivial transformations of one of two  spinor moving frame variable, $v^-_\alpha$, Eqs. (\ref{susy-v-AdS}), (\ref{susy-if=}).

Notice that this $v^-_\alpha$ is just the  spinor variable which enters the original massless superparticle action (\ref{SmM0=3D}),
which now is involved into the nAmW action as the kinetic term for the center of energy variables. On the other hand, a new feature of the action (\ref{SmM0=AdS3}) for nAmW system in AdS superspace is the explicit appearance of the Cartan  form $f^\#$ (\ref{f++:=}), constructed from the other spinor moving frame variable $v^+_\alpha$.

\section{Conclusion and outlook}

In this paper we have constructed the  action for an interacting multiple wave system  in flat and AdS  $D=3$ ${\cal N}=1$ superspaces, which we call {\it non-Alelian multiwave} or {\it nAmW}. This action is invariant under the rigid spacetime supersymmetry and local (gauge) worldline supersymmetry which acts on the Abelian center of energy variables by a deformation of the irreducible $\kappa$--symmetry transformations characteristic for the twistor--like formulation of the massless superparticle.

The action has the form (\ref{SmM0=S+AdS}) or, more explicitly,
\begin{eqnarray}
\label{SmM0=AdS3} S^{AdS}_{nAmW} & =& \int_{W^1} \rho^{\#}\, {E}^{=} + {1\over \mu^6} \int_{W^1} (\rho^{\#})^3\, \left(
tr\left(- {\bb P}D {\bb X} + {i\over 4} { \Psi} D {\Psi}  \right) + {E}^{\#} {\cal H} + i {E}^{+}  tr({\Psi}
{\bb P})\right)-\quad \nonumber \\ && -  {\zeta \over \mu^{12}}\int_{W^1}    (\rho^{\#})^5 f^{\#}\, (tr({\bb P}{\bb X}))^2
, \;
  \end{eqnarray}
where $\zeta$ is proportional to the inverse radius of AdS space, $\mu$ is a constant of dimension of mass and  ${\cal H}$ and $E^\#, E^=, E^+$ are defined by  Eqs.  (\ref{HSYM=3D}), (\ref{E++=Eu++}),(\ref{E==Eu=}) and (\ref{dEa=AdS}). In addition to coordinate functions and spinor moving frame variables (bosonic $v_\alpha^{\pm}$ constrained by (\ref{v-v+=1})), describing the center of energy motion of nAmW system, the action (\ref{SmM0=AdS3}) contains the traceless $N\times N$ matrix fields, bosonic  ${\bb X}$, fermionic ${\Psi}$ and auxiliary bosonic
${\bb P}$ and ${\bb A}=d\tau {\bb A}_\tau $, which describe the relative motion of $N$ constituents of the 3D nAmW system.

This form of the action uses the spinor moving frame formalism, which, in our opinion, makes its structure more transparent and also allows to clarify its relation with locally supersymmetric generalization of the SYM Lagrangian form   (\ref{LSYM=3D}). However, in Sec. 2 we have also presented the action in terms of unconstrained spinors (see Eqs. (\ref{SnAW:=}), (\ref{L0:=}) and (\ref{LSUN:=})).

The flat superspace limit ($\zeta=0$) of (\ref{SmM0=AdS3}) describes a simplified counterpart of the action for
11D multiple M0-brane (mM$0$) system \cite{mM0action}  (see (\ref{SmM0=})), which is known for the case of
flat target superspace only. Thus our 3D nAmW system   can be used as a toy model  to develop the tools to be applied in studying multiple M$0$-branes, in particular the methods to attack the problem of curved superspace generalization of the 11D mM$0$ action  (\ref{SmM0=}).

In this respect, the fact that we have succeed in constructing the generalization of flat superspace nAmW action to rigid but curved D=3 AdS superspace looks promising. Now one can try to use this experience to attack the problem of generalization of the mM$0$   action for the case of curved but rigid $AdS_{4(7)}\times {\bb S}^{7(4)}$ superspace(s). Such an action may be interesting in the perspective of AdS/CFT correspondence. To search for mM$0$ action in generic 11D supergravity superspace is a more complicated problem and, as a first stage it might be useful to construct the generalization of our nAmW action (\ref{SmM0=3D}) to the arbitrary curved D=3 ${\cal N}=1$ superspace.

Another interesting problem is to look for the generalization of our multiwave actions for the case of higher branes. Again, it looks convenient to start from the  case of 3D systems trying to generalize our action (\ref{SmM0=AdS3}),  or its flat superspace reduction (\ref{SmM0=3D}), for the case of 3D ${\cal N}=1$ superstring.

Notice that to describe the systems of 3D Dirichlet-$p$-branes (3D mD$p$) one needs to work in the ${\cal N}=2$ $D=3$ superspace. The action for  spacetime feeling multiple D$2$-brane system in flat ${\cal N}=2$ 3D superspace was found in  \cite{Drummond:2002kg} in the frame of superembedding approach \cite{bpstv,bsv,hs2} (see \cite{Dima99} for the review and more references). Curiously, this action is not defined uniquely; it happens to be given by a sum of terms which are $\kappa$--symmetric by themselves. This lack of uniqueness can be treated as a counterpart  of the presence of an arbitrary function of matrix variables in the 10D mD$0$  action of \cite{Dima+=JHEP03}. In contrast, the 11D mM$0$ action of \cite{mM0action} is fixed uniquely by the requirement of $\kappa$--symmetry. As it was discussed in \cite{mM0action}, it is actually possible to reproduce an mD$0$ action with an arbitrary function by dimensional   reduction of the unique mM$0$ action, provided it is allowed to assume an arbitrary dependence of the momentum in 10-th, compactified direction on the matrix fields. Probably the non-uniqueness of the 3D mD$2$ action of \cite{Drummond:2002kg} can be reproduced in a similar way starting from a (still hypothetical) unique action of 4D ${\cal N}=1$ multiple membrane system.

The search for such 4D ${\cal N}=1$ multiple membrane action\footnote{The very low energy approximation for such a 4D ${\cal N}=1$ multiple membrane system, its 'conformal fixed point',  can be obtained by a kind of `matrix' dimensional reduction of the ABJM action \cite{ABJM}  or, in the case of $N=2$ and $N=\infty$, of the BLG action \cite{BLG,Bagger:2012jb}.} can begin with constructing its simpler counterpart, the 4D ${\cal N}=1$ nonAbelian multiwave system. This will be given by  the $D=4$ generalization of the present $D=3$ action and shall provide a bit more complex but still toy--model--type counterpart of the mM$0$ system, which, in distinction to our present $D=3$ model, shall contain a nontrivial potential term for two  bosonic  matrix fields.
This D=4 model is under investigation now.

{\bf Acknowledgements}. { This work was supported in part by the research grant
FPA2012-35043-C02-01 from  MINECO of Spain and by the Basque Government research group grant ITT559-10.}
%and by the UPV/EHU under the program UFI 11/55.}

\newpage

\renewcommand{\theequation}{A.\arabic{equation}}

\begin{appendix}

\section*{Appendix: Spinor moving frame and twistor--like action of the 3D massless superparticle}
\label{3D}
\setcounter{equation}{0}
In this appendix we present some details on spinor moving frame formulation of  3D massless superparticle,  (\ref{S0=3D}),
and  on the properties of  spinor moving frame variables, the constrained spinors $v_\alpha^\pm$, and related Cartan forms.

As it was discussed in sec. 2.3, introducing the St\"uckelberg field $\rho^{\#}$ and using it to write the bosonic spinor $\lambda_\alpha$ as in (\ref{l=vr}), we can present the twistor--like action (\ref{S0=3D}) in the equivalent form  (\ref{S0=3D=s2}),
\begin{eqnarray}
\label{S0=3DA} S^{3D}_0 &=& \int_{W^1} \rho^{\#} {E}^a u_a^{=} := \int_{W^1} \rho^{\#} \, v^{-} \gamma_{a}v^{-}\,
{E}^a \; , \quad
\end{eqnarray}
where
$ u_a^{=}= v^-\gamma_av^- $
is the light--like vector ($ u^{a=} u_a^{=}=0$ ) constructed as a bilinear of the bosonic spinors $v_\alpha^-$, Eq. (\ref{u--=3D=s2}).

It is convenient to introduce the second bosonic spinor,  $v_\alpha^+$, complementary to
 $v_\alpha^+$ in the sense of  normalization condition
 \begin{eqnarray}
\label{v-v+=1=A} v^{-\alpha}v_\alpha^+ := \epsilon^{\alpha\beta} v_\alpha^+ v_\beta^-=1 \; .  \qquad
\end{eqnarray}
This condition is tantamount to saying that the determinant of the 2$\times 2$ matrix $(v_\alpha^+ , v_\alpha^-)$ is equal
to unity, {\it i.e.} that this matrix takes its values in  $SL(2,\bb{R})$ group,
\begin{eqnarray}
\label{detV=1} V_\alpha^{(\beta)}:= (v_\alpha^+ , v_\alpha^-)\quad \in \quad SL(2,\bb{R}) \qquad \Leftrightarrow \qquad det
(V_\alpha^{(\beta)})=1\; \; .  \qquad
\end{eqnarray}
As far as $SL(2,\bb{R})=Spin(1,2)$, the above matrix is in (two--to-one) correspondence with some $SO(1,2)$ valued matrix,
which should describe some dynamical frame (coordinate  system) related to the worldline of the superparticle. To construct
this {\it moving frame} matrix, let us define the bilinear of the $v^+$ spinors, the light--like vector
$ u_a^{\#}= v^+\gamma_av^+ $, Eqs. (\ref{u++=3D}).  This obeys $u^{a\#}u_a^{\#}=0$ and  $u^{a=}u_a^{\#}=2$  due to (\ref{v-v+=1}) and the $D=3$ identity $\gamma^a_{\alpha\beta}
\gamma_a^{\gamma\delta}=2\delta_{\alpha} ^{[\gamma}\delta_{\beta} ^{\delta ]}$.

Finally, the product of $v^+$ and $v^-$ determines the third vector, $ u_a^{\perp}= v^+\gamma_av^- $, Eqs. (\ref{u2=3D}),
which is space--like, normalized to $-1$  and orthogonal to  both $u_a^{\#}$ and $u_a^=$. The properties of vectors composed from the two normalized bosonic spinors,
\begin{eqnarray}
\label{u=vga} u_a^{=}= v^-\gamma_av^- \; , \qquad u_a^{\#}= v^+\gamma_av^+ \; , \qquad u_a^{\perp}= v^-\gamma_av^+ \;  \qquad
\end{eqnarray}
are resumed in Eqs. (\ref{uu=3D}),
which is tantamount to saying that these vectors can be used to form an $SO(1,2)$ valued matrix,
\begin{eqnarray}
\label{UinSpin3D}
 U_a^{(b)}:= \left({1\over 2}(u_a^{\#}+u_a^{=}) ,  u_a^{\perp} , {1\over 2}(u_a^{\#}-u_a^{=}) \right)   \; \in \; SO(1,2)
 \quad
 \\ \nonumber \Leftrightarrow \qquad  U^T\eta U=\eta =diag (1,-1,-1)\, ,  \quad
\end{eqnarray}
called {\it moving frame matrix}. To justify this name let us notice that, using (\ref{l=vr}) and (\ref{u--=3D}) we can
write  Eq. (\ref{p=vgv=3D}) in the form
\begin{eqnarray}
\label{p=ru} p_a= u_a^= \rho^{\#} = U_a^{(b)} (1, 0, -1) \rho^{\#} \quad
\end{eqnarray}
which implies that $U_a^{(b)}$ describes the Lorentz transformation from  special frame where the massless superparticle
momentum has the form $p_{(b)}= \rho (\,1,\; 0,-1)$ to a generic frame under consideration.\footnote{Notice that the above
moving frame variables were called 'light--cone harmonics' in  \cite{Sok} and vector harmonic variables in \cite{B90}.}

Then the unimodular matrix $V$ in (\ref{detV=1}) is called {\it spinor moving frame matrix} because it is the double
covering of the moving frame matrix, {\it i.e.} it obeys
\begin{eqnarray}
\label{U=VgV} V\gamma^{(b)}V^T = \gamma^a U_a^{(b)}\; , \qquad V^T\tilde{\gamma}_aV = U_a^{(b)}\tilde{\gamma}{}_{(b)} \quad
\end{eqnarray}
Using the explicit representation for D=3 gamma matrices in which $\gamma^0= \tilde{\gamma}^0= I$,  $\gamma^1= -
\tilde{\gamma}^1= \tau^1$, $\gamma^2= - \tilde{\gamma}^2= \tau^3$, $\epsilon^{12}=-\epsilon_{12}=1$, we find that Eqs.
(\ref{U=VgV}) are equivalent to the set of composition relations for the moving frame vectors in terms of moving frame
spinors, which have been collected in (\ref{u--=3D}), (\ref{u++=3D}) and (\ref{u2=3D}).

In the main text  we have used  the covariant Cartan forms  of the coset $SL(2,{\bb R})/SO(1,1)$ parametrized by the spinor moving
frame variables, (\ref{f=:=}),
\begin{eqnarray}
\label{f=:=A} f^{=}= v^{\alpha -}d v_{\alpha}^{-} \; , \qquad f^{\#}= v^{\alpha +}d v_{\alpha}^{+} \; . \qquad \;   \qquad
  \end{eqnarray}
These do not enter the action of nAmW system in flat superspace (\ref{SmM0=3D}), but one of them appears explicitly in the nAmW action in AdS superspace (\ref{SmM0=S+AdS}), (\ref{SmM0=AdS3}).

The covariant derivatives of moving frame and spinor moving frame variables, constructed with the use of the connections
(\ref{3D:Om0=}), are expressed in terms of these Cartan forms,
\begin{eqnarray}
\label{Dv:=3D} && D v_{\alpha}^{-}= f^{=} v_{\alpha}^{+}\; , \qquad Dv_{\alpha}^{+}= - f^{\#} v_{\alpha}^{-}\; , \qquad
\\ \label{Du:=3D} && Du_a^{=}= u_a^{\perp}  f^{=}\; , \qquad Du_a^{\#}= - f^{\#}  u_a^{\perp}
  \; , \qquad  D u_a^{\perp} = {1\over 2} f^{=} u_a^{\#}+ {1\over 2} f^{\#}u_a^{=}\; . \qquad
  \end{eqnarray}
Also the variations of these constrained variables can be expressed through the set of 3 independent variations which can be
identified with the formal contraction of the above Cartan froms with variation symbol. The nontrivial variations are
\begin{eqnarray}
\label{vvar}  &&  \delta  v_{\alpha}^{-}= i_\delta  f^{=} v_{\alpha}^{+}\; , \qquad  \delta  v_{\alpha}^{+} =  - i_\delta
f^{\#} v_{\alpha}^{-}  \qquad  \\ &&
 \delta u_a^{=}= u_a^{\perp}  i_\delta  f^{=}\; , \qquad  \delta u_a^{\#}= - i_\delta  f^{\#}  u_a^{\perp}
  \; , \qquad   \delta   u_a^{\perp} = {1\over 2} i_\delta  f^{=} u_a^{\#}+ {1\over 2} i_\delta  f^{\#}u_a^{=}\; , \qquad
\end{eqnarray}
while the transformations with parameter $i_\delta \Omega^{(0)}$ can be easily recognized as $SO(1,1)$ local  transformation
which is the gauge symmetry of the action (\ref{SmM0=3D}).

\end{appendix}

\end{document}